%% file: main.tex
\documentclass[runningheads]{llncs}

\input{preamble}

\input{macros}
\input{figs}

\title{Efficient Loop Conditions for\\ Bounded Model Checking
  Hyperproperties\thanks{This research has been partially supported by
    the United States NSF SaTC Award 2100989, by the Madrid Regional
    Gov. Project BLOQUES-CM (S2018/TCS-4339), by Project
    PRODIGY (TED2021-132464B-I00) funded by
    MCIN/AEI/10.13039/501100011033/ and the EU
    NextGenerationEU/PRTR, and by a research grant from Nomadic Labs
    and the Tezos Foundation.}}

\begin{document}

\maketitle
\input{abstract}





\input{intro}
\input{prelim}
\input{adaptation}
\input{algos}
\input{exp}
\input{concl}

\bibliographystyle{plain}
\bibliography{bibliography}

\newpage
\appendix
\input{proofs}

\end{document}

%% file: preamble.tex
\usepackage{tabulary}
\usepackage{fancyvrb}
\usepackage{multirow}
\usepackage{array}
\usepackage{booktabs}   
\usepackage{subcaption} 
\usepackage[new]{old-arrows}
\usepackage{xspace}
\usepackage[utf8]{inputenc}
\usepackage[T1]{fontenc}
\usepackage{array}
\usepackage{tabularx}
\usepackage{makecell}
\usepackage{paralist}
\usepackage{wrapfig}
\usepackage{listings}
\usepackage{tcolorbox}
\usepackage{color, colortbl}
\usepackage{pdfpages}

\usepackage{amsmath}
\usepackage{amsmath,amssymb}
\usepackage{amsfonts}
\usepackage{ltlfonts}

\usepackage{amssymb}
\usepackage{pifont}
\usepackage{stackrel}

\usepackage{enumitem}

\usepackage{todonotes}
\usepackage{soul}

\usepackage{algcompatible}
\DeclareCaptionFormat{myformat}{#3}
\usepackage[ruled,vlined,linesnumbered]{algorithm2e}

\usepackage{tikz}
\tikzset{
	>=stealth,
	possible world/.style={circle,draw,thick,align=center},
	real world/.style={double,circle,draw,thick,align=center},
}
\usetikzlibrary{arrows.meta, positioning}
\tikzset{%
	block/.style = {draw=\boxcolor,rectangle,thick,
		minimum height=2em, text width=10em, align=center},
	line/.style = {draw=\boxcolor, line width=4pt, 
		-{Triangle[length=10pt, width=10pt]}, shorten >=2pt, shorten <=2pt},
}
\usetikzlibrary{positioning,mindmap,automata,fit,backgrounds,shapes, 
	arrows,shapes.misc,patterns}
\usetikzlibrary{decorations.pathmorphing}
\usetikzlibrary{decorations.markings}
\usetikzlibrary{calc}
\usetikzlibrary{fit, calc}

\usepackage{graphicx}
\usetikzlibrary{arrows, arrows.meta}
\usepackage{adjustbox}
\tikzset{arw/.style={>={Triangle[length=3mm,width=3mm]},line width=2mm,draw=gray}}

\usepackage{cite}
\usepackage{stmaryrd} 
\usepackage{hyperref}
\usepackage{marvosym}
\hypersetup{%
	colorlinks=true,           
	allcolors=blue!70!black,   
	pdfstartview=Fit,          
	breaklinks=true
}

\makeatletter
\def\@citecolor{blue}%
\def\@urlcolor{blue}%
\def\@linkcolor{blue}%

\def\orcidID#1{\smash{\href{http://orcid.org/#1}{\protect\raisebox{-1.25pt}{\protect\includegraphics{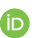}}}}}
\makeatother

\author{Tzu-Han Hsu\inst{1}\orcidID{0000-0002-6277-2765}
	\and  César Sánchez\inst{2}\orcidID{0000-0003-3927-4773}
	\and Sarai Sheinvald\inst{3}\orcidID{0000-0002-0524-7390} 
	\and \\ \Letter~Borzoo Bonakdarpour\inst{1}\orcidID{0000-0003-1800-5419}
}

\authorrunning{T.-H. Hsu et al.}

\institute{
	Michigan State University, East Lansing, MI, USA \email{\{tzuhan,borzoo\}@msu.edu}
	\and
	IMDEA Software Institute, Madrid, Spain
	\email{cesar.sanchez@imdea.org}
	\and
	Dept. of Software Engineering, Braude College, Israel
	\email{sarai@braude.ac.il}
}

%% file: macros.tex
\definecolor{gray}{rgb}{0.5,0.5,0.5}
\definecolor{darkgreen}{rgb}{0,0.6,0}
\definecolor{darkblue}{rgb}{0.2, 0.2, 0.6}
\definecolor{niceblue}{rgb}{0.16, 0.32, 0.75}
\definecolor{niceblack}{rgb}{0.0, 0.18, 0.39}
\definecolor{prettyred}{rgb}{1.0, 0.13, 0.32}
\definecolor{prettypink}{rgb}{0.98, 0.38, 0.5}
\definecolor{prettyblue}{rgb}{0.0, 0.30, 1.0}
\definecolor{prettygreen}{rgb}{0.0, 0.5, 0.0}
\definecolor{prettyorange}{rgb}{1.0, 0.33, 0.0}
\definecolor{prettypurple}{rgb}{0.6, 0.4, 0.8}
\definecolor{prettygreen}{rgb}{0.01, 0.75, 0.24}
\definecolor{prettyyellow}{rgb}{0.93, 0.53, 0.18}

\newcommand{\KWD}[1]{\mathit{#1}}

\usepackage{amsmath}
\newcommand{\OR}{\vee}
\newcommand{\AND}{\wedge}
\newcommand{\always}{\G}
\newcommand{\Always}{\G}
\newcommand{\Event}{\F}
\newcommand{\Next}{\X}
\newcommand{\G}{\LTLsquare}
\newcommand{\F}{\LTLdiamond}
\newcommand{\X}{\LTLcircle}

\newcommand{\until}{\mathbin{\mathcal{U}}}
\newcommand{\release}{\mathbin{\mathcal{R}}}

\newcommand{\DefinedAs}{\,\stackrel{\text{def}}{=}\,}
\newcommand{\code}[1]{\textsf{\small #1}\xspace}

\newcommand{\hltl}{\code{HyperLTL}}
\newcommand{\HyperLTL}{\hltl}

\newcommand{\LTL}{\code{LTL}}

\newcommand{\tru}{\mathsf{true}}

\newcommand{\States}{S}
\newcommand{\state}{s}

\newcommand{\trans}{\delta}

\newcommand{\krip}{K}
\newcommand{\Kr}{K}
\newcommand{\KrFamily}{\mathcal{K}}

\newcommand{\Tr}{\mathcal{T}}

\newcommand{\ktuple}{\langle S, S^\init, \trans, \AP, L \rangle}

\newcommand{\lang}{\mathcal{L}}
\newcommand{\init}{0}

\newcommand{\Traces}{\textit{Traces}}
\newcommand{\zplus}{\mathbb{Z}_{\geq 0}}
\newcommand{\V}{\mathcal{V}}
\newcommand{\Vars}{\textit{Vars}}
\newcommand{\tupleof}[1]{\langle#1\rangle}
\newcommand\modified[1]{#1}

\newcommand{\quant}{\mathbb{Q}}
\newcommand{\HyperQube}{\code{HyperQB}\xspace}

\def\HiLiy{\leavevmode\rlap{\hbox to \hsize{\color{yellow!50}\leaders\hrule height .8\baselineskip depth .5ex\hfill}}}
\def\HiLir{\leavevmode\rlap{\hbox to \hsize{\color{red!50}\leaders\hrule height .8\baselineskip depth .5ex\hfill}}}
\def\HiLib{\leavevmode\rlap{\hbox to \hsize{\color{blue!50}\leaders\hrule height .8\baselineskip depth .5ex\hfill}}}

\newcommand{\rh}{\textit{\rh}}
\newcommand{\RH}{\textit{\RH}}
\newcommand{\lh}{\textit{\lh}}
\newcommand{\LH}{\textit{\LH}}

\newcommand{\emp}{\em\color{red}}
\newcommand{\boxcolor}{orange}

\newcommand{\sysA}{{K}_{A}}
\newcommand{\sysB}{{K}_{B}}

\newcommand{\katuple}{\langle S_A, S_{A}^\init, \trans_A, \AP_A,  L_A \rangle}

\newcommand{\alphabet}{\mathrm{\Sigma}}
\newcommand{\AP}{\mathsf{AP}}

\newcommand{\bridge}{\textit{B}}

\newcommand{\naturals}{\mathcal{N}}
\newcommand\comp[1]{#1}
\newcommand{\tuple}[2]{ #1, #2 }
\newcommand{\f}{\varphi}
\newcommand{\g}{\psi}

\newcommand{\simrelation}{{R}}

\newcommand{\commentout}[1]{}

\newcommand{\aesim}{\code{SIM$_{\text{AE}}$}}
\newcommand{\easim}{\code{SIM$_{\text{EA}}$}}
\newcommand{\unrolling}{\code{BMC}}

\newcommand{\pred}{\code{Pred}}

\newcommand{\AAA}{P}
\newcommand{\BBB}{Q}

\newcommand{\Mp}{K_\AAA}

\newcommand{\Mq}{K_\BBB}

\newcommand{\transit}[1]{\delta_#1}

\newcommand{\simulate}{{T}}

\newcommand{\msmall}{\texttt{small}}
\newcommand{\mbig}{\texttt{big}}

\newcommand{\mypar}[1]{\noindent\textbf{\textup{#1.}}}

\newcommand{\SIM}{\KWD{sim}}
\newcommand{\Into}{\mathrel{\rightarrow}}
\newcommand{\Iff}{\mathrel{\leftrightarrow}}
\renewcommand{\And}{\mathrel{\wedge}}

\newcommand{\SUCC}{\KWD{succ}}

\newcommand{\sender}{\texttt{sender}}
\newcommand{\receiver}{\texttt{receiver}}
\newcommand{\send}{\textit{send}}
\newcommand{\receive}{\textit{receive}}
\newcommand{\wait}{\textit{wait}}

\newcommand{\varphiEA}{\ensuremath{\varphi_{\textsf{EA}}}}
\newcommand{\varphiEAnk}{\ensuremath{\varphiEA}^{n,k}}
\newcommand{\varphiAE}{\ensuremath{\varphi_{\textsf{AE}}}}
\newcommand{\varphiAEnk}{\ensuremath{\varphiAE}^{n,k}}

\newcommand{\Pre}{\textit{Pre}}
\newcommand{\Post}{\KWD{Post}}
\newcommand{\annot}{\KWD{annot}}

\newcommand{\nbw}{A}
\newcommand{\pstate}[1]{s_p(#1)}
\newcommand{\qstate}[1]{s_q(#1)}
\newcommand{\ptrace}{t_p}
\newcommand{\qtrace}{t_q}

\newcommand{\tacas}{\color{black}}
\newcommand{\M}{\Kr}

%% file: figs.tex
\newcommand{\KRIPKEA}
{
\begin{tikzpicture}[-,>=stealth,shorten >=.5pt,auto,node distance=2cm, semithick, initial text={}]
	\node[initial, state] [text width=1em, text centered, minimum  height=2.0em](0) at (0,0) {\hspace*{-0.8mm}$\{ \}$};
	\node[above left = 0.005 cm and -0.1 cm of 0](label){\Large$s_1$};
	\node[state][text width=1em, text centered, minimum 
	height=2.5em] (1) at (2,0) {\hspace*{-0.8mm}$\{ \}$};
	\node [above left = 0.005 cm and -0.1 cm of 1](label){\Large$s_{2}$};
	\node[state][text width=1em, text centered, minimum 
	height=2.5em] (2) at (4,1){\hspace*{-0.8mm}$\{a\}$};
	\node [above left = 0.005 cm and -0.1 cm of 2](label){\Large$s_{3}$};
	\node[state][text width=1em, text centered, minimum 
	height=2.5em] (3) at (4,-1){\hspace*{-0.8mm}$\{ \}$};
	\node [above left = 0.005 cm and -0.1 cm of 3](label){\Large$s_{4}$};
	\draw[->]  
	(0)	edge  node (01 label) {} (1)
	(1) edge node (12 label) {} (2)
	(1) edge node (03 label) {} (3)
	(2) edge [bend right] node (03 label) {} (0)
	(3) edge [bend left] node (03 label) {} (0);
\end{tikzpicture}
}

\newcommand{\KRIPKEB}
{
			\begin{tikzpicture}[-,>=stealth,shorten >=.5pt,auto,node distance=2cm, semithick, initial text={}]
	\node[initial, state] [text width=1em, text centered, minimum  height=2.0em](0) at (0,0) {\hspace*{-0.8mm}$\{ \}$};
	\node[above left = 0.005 cm and -0.1 cm of 0](label){\Large$q_1$};
	\node[state][text width=1em, text centered, minimum 
	height=2.5em] (1) at (2,1) {\hspace*{-0.8mm}$\{ \}$};
	\node [above left = 0.005 cm and -0.1 cm of 1](label){\Large$q_{2}$};
	\node[state][text width=1em, text centered, minimum 
	height=2.5em] (2) at (2,-1){\hspace*{-0.8mm}$\{ \}$};
	\node [above left = 0.005 cm and -0.1 cm of 2](label){\Large$q_{3}$};
	\node[state][text width=1em, text centered, minimum 
	height=2.5em] (3) at (4,1){\hspace*{-0.8mm}$\{a\}$};
	\node [above left = 0.005 cm and -0.1 cm of 3](label){\Large$q_{4}$};
	\node[state][text width=1em, text centered, minimum 
	height=2.5em] (4) at (4,-1){\hspace*{-0.8mm}$\{ \}$};
	\node [above left = 0.005 cm and -0.1 cm of 4](label){\Large$q_{5}$};

	\draw[->]  
	(0)	edge  node (01 label) {} (1)
	(0)	edge  node (01 label) {} (2)
	(1) edge node (12 label) {} (3)
	(2) edge node (03 label) {} (4)
	(3) edge [bend right=85] node (03 label) {} (0)
	(4) edge [bend left=60] node (03 label) {} (0);
\end{tikzpicture}
}

%% file: abstract.tex
\begin{abstract}
  Bounded model checking (BMC) is an effective technique for hunting bugs by incrementally
  exploring the state space of a system.
  To reason about infinite traces through a finite structure and
  to ultimately obtain completeness, BMC incorporates
  \emph{loop conditions} that revisit previously observed states.
  This paper focuses on developing loop conditions for BMC of \HyperLTL -- a temporal logic for 
  hyperproperties that allows expressing important policies for security and consistency in
  concurrent systems, etc.
  %
  %
  Loop conditions for \HyperLTL are more complicated than for \LTL, as
  different traces may loop inconsistently in unrelated moments.
  Existing BMC approaches for \HyperLTL only considered linear
  unrollings without any looping capability, which precludes both
  finding small infinite traces and obtaining a complete technique.
  {\tacas
  We investigate loop conditions for \HyperLTL BMC, for \HyperLTL
  formulas that contain up to one quantifier alternation.
  }
%
  %
  We first present a general complete automata-based technique which
  is based on bounds of maximum unrollings.
  Then, we introduce alternative simulation-based algorithms that
  allow exploiting short loops effectively, generating SAT queries
  whose satisfiability guarantees the outcome of the original model
  checking problem.
  We also report empirical evaluation of the {\tacas prototype
    implementation of our BMC techniques using \code{Z3py}.}
\end{abstract}

%% file: intro.tex
\section{Introduction}
\label{sec:intro}

Hyperproperties~\cite{cs10} have been getting increasing attention due
to their power to reason about important specifications such as
information-flow security policies that require reasoning about the
interrelation among different execution traces.
%
%
\HyperLTL~\cite{clarkson14temporal} is an extension of the linear-time temporal logic 
\LTL~\cite{pnueli77temporal} that allows
quantification over traces; hence, capable of describing
hyperproperties.
For example, the security policy {\em observational determinism} can
be specified as \HyperLTL formula:
\(
 \forall \pi.\forall \pi'.(o_{\pi} \leftrightarrow o{_{\pi'}})
\, \mathcal{W} \, \neg (i_{\pi} \leftrightarrow i{_{\pi'}}), 
\)
which specifies that for every pair of traces $\pi$ and $\pi'$, if
they agree on the secret input $i$, then their public output $o$ must
also be observed the same (here `$\mathcal{W}$' denotes the weak until
operator).

Several works~\cite{frs15,coenen19verifying} have studied model
checking techniques for \HyperLTL specifications, which typically reduce
this problem to \LTL model checking queries of modified systems.
More recently,~\cite{hsu21bounded} proposed a QBF-based algorithm for
the direct application of bounded model checking (BMC)~\cite{cbrz01}
to \HyperLTL, and successfully provided a push-button solution to
verify or falsify \HyperLTL formulas with an arbitrary number of quantifier
alternations.
However, unlike the classic BMC for \LTL, which included the
so-called {\em loop conditions}, the algorithm in~\cite{hsu21bounded}
is limited to (non-looping) linear exploration of paths.
The reason is that extending path exploration to include loops when
dealing with multiple paths simultaneously is not straightforward.
For example, consider the \HyperLTL formula
\( 
\varphi_1 = \forall {\pi}. \exists {\pi'}. ~\G(a_{\pi}\rightarrow b_{\pi'}) 
\) 
and a pair of Kripke structures $K_1$ and $K_2$ as
follows:
\begin{center}
   \vspace{-2em}
	\begin{tabularx}{\textwidth}{m{3em}@{}X@{\hspace{5em}}m{3em}@{}X} 
$K_1$&		\parbox[c]{\hsize}{\scalebox{0.6}{\KRIPKEA}} & $K_2$ & \parbox[c]{\hsize}{\scalebox{0.6}{\KRIPKEB}} 
	\end{tabularx}
\end{center}

\vspace*{-1em}
\noindent Assume trace $\pi$ ranges over $K_1$ and  trace $\pi'$ ranges over $K_2$.
Proving $\langle K_1, K_2 \rangle \not\models \varphi_1$ can be achieved by finding a finite  
counterexample (i.e., path $s_1s_2s_3$ from $K_1$).
Now, consider
\(
\varphi_2 = \forall {\pi}. \exists {\pi'}. ~\G(a_{\pi}\leftrightarrow a_{\pi'}).
\)
It is easy to see that $\langle K_1, K_2 \rangle\models \varphi_2$. 
However, to prove $\langle K_1, K_2 \rangle \models \varphi_2$, one
has to show the absence of counterexamples in infinite paths, which
is impossible with model unrolling in finite steps as proposed
in~\cite{hsu21bounded}.

In this paper, we propose efficient loop conditions for BMC of hyperproperties.
First, using an automata-based method, we show that lasso-shaped
traces are sufficient to prove infinite behaviors of traces within finite
exploration.
However, this technique requires an unrolling bound that renders it
impractical.
Instead, our efficient algorithms are based on the notion of {\em
  simulation}~\cite{Pnu85} between two systems.
Simulation is an important tool in verification, as it is used
for abstraction, and preserves \code{ACTL$^{*}$} properties
\cite{BBLS92,GL94}.
As opposed to more complex properties such as language containment,
simulation is a more local property and is easier to check. 
The main contribution of this paper is the introduction of practical algorithms that
achieve the exploration of infinite paths following a simulation-based
approach that is capable of relating the states of multiple models
with correct successor relations.

We present two different variants of simulation, \easim and \aesim,
allowing to check the satisfaction of $\exists\forall$ and $\forall\exists$ hyperproperties, 
respectively. These notions circumvent the need to 
boundlessly unroll traces in both structures and synchronize them.
For \aesim, in order to resolve non-determinism in the first model, 
we also present a third variant, where we enhance \aesim
by using {\em prophecy variables}~\cite{al91,beutner22prophecy}.
Prophecy variables allow us to handle
cases in which $\forall\exists$ hyperproperties hold despite the lack of a direct
simulation.
With our simulation-based approach, one can capture infinite behaviors of traces 
with finite exploration in a simple and concise way.
Furthermore, our BMC approach not only model-checks the systems for 
hyperproperties, but also does so in a way that finds {\em minimal} witnesses to
the simulation (i.e., by partially exploring the existentially quantified model), 
which we will further demonstrate in our empirical evaluation. 
%

\begin{wraptable}{r}{.638 \textwidth}
	\vspace*{-6mm}
	\renewcommand{\arraystretch}{1.2}
	\centering 
	\scalebox{1}{
		\input{table_complete}
	}
	\caption{
		Eight categories of \hltl formulas with different forms of quantifiers, sizes of models, and different temporal operators. 
	}
	\label{tab:category}
	\vspace*{-5mm}	
\end{wraptable}
%

We also design algorithms that generate SAT formulas for each variant 
(i.e., \easim, \aesim, and \aesim with prophecies), where the
satisfiability of formulas implies the model checking outcome.
We also investigate the practical cases of models with different sizes
leading to the eight categories in Table~\ref{tab:category}.
For example, the first row indicates the category of verifying two models
of different sizes with the fragment that only allows $\forall\exists$ quantifiers and $\G$ (i.e., {\em 
globally} temporal operator);
$\forall_\msmall \exists_\mbig$ means that the first model is
relatively smaller than the second model, and the positive
outcome ($\models\forall\exists\G\varphi$) can be proved by our
simulation-based technique \aesim, while the negative outcome
($\not\models\forall\exists\G\varphi$) can be easily checked using
non-looping unrolling (i.e.,~\cite{hsu21bounded}).
We will show that in certain cases, one can verify a $\G$ formula
without exploring the entire state space of the $\mbig$ model to 
achieve efficiency.

%

{\tacas We have implemented our
  algorithms\addtocounter{footnote}{-3}\footnote{Available at:
    \url{https://github.com/TART-MSU/loop_condition_tacas23}} using
  \code{Z3py}, the \code{Z3}~\cite{dmb08} API in python.}
%
We demonstrate the
efficiency of our algorithm exploring a subset of the state space for
the larger {\tacas (i.e., \texttt{big})} model.
We evaluate the applicability and efficiency with cases including
conformance checking for distributed protocol synthesis, model
translation, and path planning problems.
In summary, we make the following contributions: (1) a bounded model
checking algorithm for hyperproperties with loop conditions, (2)
three different practical algorithms: \easim, \aesim, and \aesim with
prophecies, and (3) a demonstration of the efficiency and
applicability by case studies that cover through all eight different
categories of \HyperLTL formulas (see Table~\ref{tab:category}).

\mypar{Related Work}
Hyperproperties were first introduced by Clarkson and
Schneider~\cite{cs10}.
\HyperLTL was introduced as a temporal logic for hyperproperties
in~\cite{clarkson14temporal}.
The first algorithms for model checking \HyperLTL were introduced
in~\cite{frs15} using alternating automata.
Automated reasoning about \hltl specifications has received attention in many aspects, including static
verification~\cite{frs15,fmsz17,fht18,cfst19} and
monitoring~\cite{ab16,fhst19,bsb17,bss18,fhst18,sssb19,hst19}.
This includes tools support, such as \code{MCHyper}~\cite{frs15,cfst19}
for model checking, \code{EAHyper}~\cite{fhs17} and
\code{MGHyper}~\cite{fhh18} for satisfiability checking, and
\code{RVHyper}~\cite{fhst18} for runtime monitoring.
%
%
{\tacas
However, the aforementioned tools are either limited to \hltl formulas
without quantifier alternations, or requiring additional inputs from the user (e.g., manually added strategies~\cite{cfst19}). 
}

Recently, this difficulty of alternating formulas was tackled by the
bounded model checker \code{HyperQB}~\cite{hsu21bounded} using QBF
solving.
However, \code{HyperQB} lacks loop conditions to capture early
infinite traces in finite exploration.
In this paper, we develop simulation-based algorithms to overcome this limitation.
%
There are alternative approaches to reason about infinite traces, like reasoning about strategies to deal with $\forall\exists$
formulas~\cite{coenen19verifying}, whose completeness can be obtained
by generating a set of prophecy variables~\cite{beutner2022prophecy}.
In this work, we capture infinite traces in BMC approach using
simulation.
We also build an applicable prototype for model-check \hltl formulas
with models that contain loops.


%% file: table_complete.tex
	\begin{tabular}[t]{|p{1.8cm} | p{2.8cm}  | p{2.7cm} |}
		\hline
		~~~~{\bf Case}  &  ~~~~~~$\varphi$ with $\G$~~~~  & ~~~~$\neg \varphi$ with $\F$~~~~ \\ [0.5ex]
		\hline 
		\hline
		$\forall_\msmall~ \exists_\mbig$  
		& \thead[l]{\hspace*{.6mm}\aesim~$ \rightarrow\hspace*{1mm}\models \forall\exists \G\varphi$ \\ 
		} 
		& \thead[l]{\hspace*{.6mm}\unrolling ~$ \rightarrow\hspace*{1mm}\not\models \forall\exists \G\varphi$ \\  
		}  \\
		\hline
		$\forall_\mbig~~~ \exists_\msmall$ 
		& \thead[l]{\hspace*{.6mm}\aesim ~$ \rightarrow\hspace*{1mm}\models \forall\exists \G\varphi$ \\ 
		}  
		& \thead[l]{\hspace*{.6mm}\unrolling ~$ \rightarrow\hspace*{1mm}\not\models \forall\exists \G\varphi$ \\ 
		} \\
		\hline
		\hline
		$\exists_\msmall~ \forall_\mbig$ 
		& \thead[l]{\hspace*{.6mm}\easim ~$ \rightarrow\hspace*{1mm}\models \exists\forall \G\varphi$  \\ 
		}  
		& \thead[l]{\hspace*{.6mm}\unrolling ~$ \rightarrow\hspace*{1mm}\not\models \exists\forall \G\varphi$ \\ 
		} \\
		\hline
		$\exists_\mbig~~~ \forall_\msmall$ 
		& \thead[l]{\hspace*{.6mm}\easim ~$ \rightarrow\hspace*{1mm}\models \exists\forall \G\varphi$ \\
		}  
		& \thead[l]{\hspace*{.6mm}\unrolling ~$ \rightarrow\hspace*{1mm}\not\models \exists\forall \G\varphi$ \\
		} \\
		\hline
	\end{tabular}

%% file: prelim.tex
\section{Preliminaries}

\mypar{\textup{Kripke structures}}
A {\em Kripke structure} $\krip$ is a tuple $\ktuple$, where $S$ is a
set of {\em states}, $S^\init \subseteq S$ is a set of {\em initial
  states}, $\trans\subseteq S\times S$ is a total {\em transition
  relation}, and $L:S\rightarrow 2^{\AP}$ is a {\em labeling
  function}, which labels {\tacas states $s \in S$} with a subset of
atomic propositions in $\AP$ that hold in $s$.
{\tacas 
%
A \emph{path} of $\krip$ is an infinite sequence of states
$\state(0)\state(1)\cdots \in \States^\omega$, such that 
$\state(0) \in \States^\init$, and
$(\state(i), \state({i+1})) \in \trans$, for all $i \geq 0$.
A {\em loop} in $\krip$ is a finite path 
$\state(n)\state(n+1)\cdots \state(\ell)$, for some $0 \leq n \leq \ell$, such that
$(\state(i), \state({i+1})) \in \trans$, 
for all $n \leq i < \ell$, and
$(\state(\ell), \state(n)) \in \trans$. 
Note that $n=\ell$ indicates a {\em self-loop} on a state.
A \emph{trace} of $\krip$ is a trace
$t(0)t(1)t(2) \cdots \in \alphabet^\omega$, such that there exists a
path $\state(0)\state(1)\cdots \in \States^\omega$ with
$t(i) = L(\state(i))$ for all $i\geq 0$.
We denote by $\Traces(\krip, \state)$ the set of all traces of $\krip$
with paths that start in state $\state \in \States$.
We use $\Traces(\krip)$ as a shorthand for
$\bigcup_{s \in \States^{\init}}\Traces(\krip,\state)$, and
$\lang(\Kr)$ as the shorthand for $\Traces(\krip)$.
}

\vspace{0.5em}
\mypar{\textup{Simulation relations}}
Let $\sysA = \katuple$ and $\sysB = \langle S_B, S_{B}^\init, \trans_B,$ $\AP_B, L_B \rangle$ be two Kripke structures. 
A {\em simulation relation} $\simrelation$ from $\sysA$ to $\sysB$ is a relation $\simrelation\subseteq S_A\times S_B$ that meets the following conditions:
\begin{compactenum}
    \item For every $s_A\in S_{A}^\init$ there exists $s_B\in S_{A}^\init$ such that $(s_A,s_B)\in \simrelation$.
    \item For every $(s_A,s_B)\in \simrelation$, it holds that $L_A(s_A)=L_B(s_B)$.
    \item For every $(s_A,s_B)\in \simrelation$, for every $(s_A, s'_A)\in \trans_A$, there exists 
    $(s_B,s'_B)\in \trans_B$ such that $(s'_A,s'_B)\in \simrelation$.
    \end{compactenum}

\vspace{0.5em}
\input{prelim-cesar}

\input{prelim-sarai}

    

%% file: prelim-cesar.tex

\mypar{The Temporal Logic HyperLTL}
\HyperLTL~\cite{clarkson14temporal} is an extension of the linear-time
temporal logic (\LTL) for hyperproperties.
The syntax of \HyperLTL formulas is defined inductively by the
following grammar:
\begin{equation*}
\begin{aligned} & \varphi ::= \exists \pi . \varphi \mid \forall
\pi. \varphi \mid \phi \\ & \phi ::= \tru \mid a_\pi \mid \lnot \phi
\mid \phi \OR \phi \mid \phi \AND \phi \mid \phi \until \, \phi \mid
\phi \release \, \phi \mid \X \phi
    \end{aligned}
\end{equation*}
where $a \in \AP$ is an atomic proposition and $\pi$ is a {\em trace
variable} from an infinite supply of variables $\V$.
The Boolean connectives $\neg$, $\OR$, and $\AND$ have the usual
meaning, $\until$ is the temporal \emph{until} operator, $\release$ is
the temporal \emph{release} operator, and $\X$ is the temporal
\emph{next} operator.
We also consider other derived Boolean connectives, such as
$\rightarrow$ and $\leftrightarrow$, and the derived temporal
operators \emph{eventually} $\F\varphi\equiv \tru\,\until\varphi$ and
\emph{globally} $\always\varphi\equiv\neg\F\neg\varphi$.
%
%
A formula is {\em closed} (i.e., a {\em sentence}) if all trace
variables used in the formula are quantified.
We assume, without loss of generality, that {\tacas no trace variable is quantified
twice.}
We use $\Vars(\varphi)$ for {\tacas the set of trace variables} used in formula
$\varphi$.

\paragraph{Semantics.}
An interpretation $\Tr=\tupleof{T_\pi}_{\pi\in\Vars(\varphi)}$ of a
formula $\varphi$ consists of a tuple of sets of traces, with one set
$T_\pi$ per trace variable $\pi$ in $\Vars(\varphi)$, denoting the set
of traces that $\pi$ ranges over.
Note that we allow quantifiers to range over different models, which
is called the {\em multi-model
semantics}~\cite{goudsmid21compositional,hsu21bounded}\footnote{In terms
  of the model checking problem, multi-model and (the conventional)
  single-model semantics where all paths are assigned traces from the
  same Kripke structure~\cite{clarkson14temporal} are equivalent
  (see~\cite{goudsmid21compositional,hsu21bounded}).}.
%
%
That is, each set of traces comes from a Kripke structure and we use
{\tacas $\KrFamily=\tupleof{K_\pi}_{\pi\in\Vars(\varphi)}$ to denote a {\em family}
of Kripke structures,} so $T_\pi=\Traces(K_\pi)$ is the traces that
$\pi$ can range over, which comes from {\tacas $K_\pi \in \KrFamily$.}
{\tacas Abusing notation, we write $\Tr=\Traces(\KrFamily)$.}
%

The semantics of \HyperLTL is defined with respect to a trace
assignment, which is a partial map~$\Pi \colon \Vars(\varphi)
\rightharpoonup\alphabet^\omega$.
The assignment with the empty domain is denoted by $\Pi_\emptyset$.
Given a trace assignment~$\Pi$, a trace variable~$\pi$, and a concrete
trace~$t \in \alphabet^\omega$, we denote by $\Pi[\pi \rightarrow t]$
the assignment that coincides with $\Pi$ everywhere but at $\pi$,
which is mapped to trace $t$.
The satisfaction of a \HyperLTL formula $\varphi$ is a binary relation
$\models$ that associates a formula to the models $(\Tr,\Pi,i)$ where
$i \in \zplus$ is a pointer that indicates the current evaluating
position.
The semantics is defined as follows:
\[
  \begin{array}{ll@{\hspace{1.4em}}c@{\hspace{1.4em}}l} (\Tr, \Pi,0)
    &\models \exists \pi.\ \modified{\psi} & \text{iff} & \text{ there
      is a } t \in T_\pi, \text{ such that } (\Tr,\Pi[\pi \rightarrow{}t],0) \models \psi,\\
    (\Tr, \Pi,0) &\models \forall \pi.\
    \modified{\psi} & \text{iff} & \text{ for all } t \in T_\pi,
                                   \text{ such that } (\Tr,\Pi[\pi\rightarrow t],0) \models \psi,\\
    (\Tr, \Pi,i) &\models \tru && \\
    (\Tr, \Pi,i) &\models a_\pi &\text{iff} & a \in \Pi(\pi)(i),\\ (\Tr, \Pi,i) &\models \neg \psi  & \text{iff} & (\Tr, \Pi,i) \not\models \psi
    (\Tr, \Pi,i) \\
    (\Tr, \Pi,i)
    &\models \psi_1 \OR \psi_2 & \text{iff} & (\Tr, \Pi,i) \models
    \psi_1\text{ or } (\Tr, \Pi,i) \models \psi_2,\\
%
%
    %
(\Tr, \Pi,i)
    &\models \psi_1 \AND \psi_2 & \text{iff} & (\Tr, \Pi,i) \models
    \psi_1 \text{ and } (\Tr, \Pi,i) \models \psi_2,\\    
 (\Tr, \Pi,i)
    &\models \X \psi & \mbox{iff} & (\Tr,\Pi,i+1)\models\psi,\\ (\Tr,
    \Pi,i) &\models \psi_1 \until \psi_2 & \text{iff} & \text{there is
      a } j \ge i \text{ for which } (\Tr,\Pi,j) \models \psi_2 \text{
      and } \\ &&& \hspace{1em} \text{for all } k \in [i, j),
    (\Tr,\Pi,k)\models \psi_1,\\ (\Tr, \Pi,i) &\models \psi_1 \release
    \psi_2 & \text{iff} & \text{either for all } j \geq i,\;
    (\Tr,\Pi,j) \models \psi_2 \text{, or, } \\ &&& \hspace{1em}
    \text{for some } j\geq i, (\Tr,\Pi,j)\models \psi_1 \text{ and }\\
    &&& \hspace{1em} \text{for all } k\in [i, j]:
    (\Tr,\Pi,k)\models\psi_2.
  \end{array}
\]
We say that an interpretation $\Tr$ satisfies a sentence~$\varphi$,
denoted by $\Tr \models \varphi$, if $(\Tr, \Pi_\emptyset,0) \models
\varphi$.
{\tacas
We say that a family of Kripke structures $\KrFamily$ satisfies a
sentence~$\varphi$, denoted by $\KrFamily \models \varphi$, if 
$\langle\Traces(\krip_\pi)\rangle_{\pi \in \Vars(\varphi)} \models \varphi$.
When the same Kripke structure $K$ is used for all path variables we
write $K\models\varphi$.
}

%


%% file: prelim-sarai.tex
\begin{definition}
	\label{def:nba}
	{\em
	A {\em nondeterministic B{\"u}chi automaton} (NBW) 
	is a tuple \linebreak 
	{\tacas $\nbw = \langle \Sigma,Q,Q_0,\delta,F \rangle$,} 
	where $\Sigma$ is an {\em alphabet}, $Q$ is a 
	nonempty finite set of {\em states}, $Q_0\subseteq Q$ is a set of {\em initial 
		states}, $F\subseteq Q$ is a set of {\em accepting states}, and 
	$\delta\subseteq Q\times\Sigma\times Q$ is a {\em transition relation}.  
	}
\end{definition}
%
Given an infinite word $w=\sigma_1\sigma_2\cdots$ over $\Sigma$, a 
{\em run of $\nbw$ on $w$} is an infinite sequence of states $r = (q_0,q_1,\ldots)$, such 
that $q_0\in Q_0$, and $(q_{i-1},\sigma_i, q_i)\in \delta$ for every $i>0$.
The run is {\em accepting} if $r$ visits some state in $F$ infinitely often.
We say that $\nbw$ {\em accepts} $w$ if there exists an accepting run of $\nbw$ on $w$. 
The {\em language} of $\nbw$, denoted $\lang(\nbw)$, is the set of all infinite words accepted by 
$\nbw$. 
An NBW $\nbw$ is called a {\em safety} NBW if all of its states are accepting.
Every safety \LTL formula $\g$ can be translated into a safety NBW over $2^{\AP}$ such that 
$\lang(\nbw)$ is the set of all traces over $\AP$ that satisfy $\g$~\cite{kv99}.

%% file: adaptation.tex
\section{Adaptation of BMC to HyperLTL on Infinite Traces}
\label{sec:adaptation}

There are two main obstacles in extending the BMC approach of~\cite{hsu21bounded} to handle
infinite traces.
First, a trace may have an irregular behavior. Second, even traces
whose behavior is regular, that is, lasso shaped, are hard to
synchronize, since the length of their respective prefixes and lassos
need not to be equal.
For the latter issue, synchronizing two traces whose {\tacas prefixes} and lassos
are of lengths $p_1,p_2$ and $l_1,l_2$, respectively, is equivalent to
coordinating the same two traces, when defining both their prefixes to
be of length {\tacas $\max\{p_1,p_2\}$}, and their lassos to be of length
{\tacas $\mathrm{lcm}\{l_1,l_2\}$, where `$\mathrm{lcm}$' stands for 
	`least common multiple'.}
%
%
As for the former challenge, we show that restricting the exploration
of traces in the models to only consider lasso traces is sound.
That is, considering only lasso-shaped traces is equivalent to
considering the entire trace set of the models.


{\tacas
Let $\Kr = \ktuple$ be a Kripke structure. 
A {\em lasso path} of $\Kr$ is a path
$s(0)s(1)\ldots s(\ell)$ such that $(s(\ell), s(n)) \in \delta$ for some {$0 \leq n <\ell$}. 
%
This path induces a {\em lasso trace} (or simply, a {\em lasso})
$L(s_0)\dots L(s_{n-1})~ (L(s_n)\dots L(s_{\ell}))^\omega$.
%
Let $\tupleof{\Kr_1,\ldots, \Kr_k}$ be a multi-model.
We denote the set of lasso traces of $\Kr_i$ by $C_i$ for all $1 \leq i \leq k$, 
and {we use $\lang{(C_i)}$ as the shorthand for the set of lasso traces of $\Kr_i$.}
}

\newcounter{thm-lassos}
\setcounter{thm-lassos}{\value{theorem}}

\begin{theorem}
  \label{thm:lassos}
	Let $\KrFamily = \tupleof{\Kr_1,\ldots, \Kr_k}$ be a multi-model, and let 
	$\f = \quant_1 \pi_1. \cdots\quant_k \\\pi_k.\g$ be a \HyperLTL formula, both over $\AP$,  then $\KrFamily \models  \f$ iff $\tupleof{C_1,\ldots, C_k}\models \f$.
\end{theorem}

\begin{proof}(sketch) 
For an \LTL formula $\g$ over $\AP\times \{\pi_i\}_{i=1}^{k}$, 
we denote  the translation of $\g$ to an NBW over 
$2^{\AP\times \{\pi_i\}_{i=1}^{k}}$ by $\nbw_\g$~\cite{vw86}. 
Given $\alpha = \quant_1 \pi_1\cdots\quant_k \pi_k$, 
where $\quant_i\in\{\exists, \forall\}$, 
we define the satisfaction of 
$A_\g$ by $\KrFamily$ w.r.t. $\alpha$, 
denoted $\KrFamily \models (\alpha$, $\nbw_\g$),  in the natural way: 
$\exists\pi_i$ corresponds to the existence of 
a path assigned to $\pi_i$ in $\Kr_i$, and dually for  $\forall\pi_i$.
Then, $\KrFamily \models (\alpha, \nbw_\g)$ iff the various $k$-assignments of traces of $\KrFamily$ to $\{\pi_i\}_{i=1}^{k}$ according to $\alpha$ are accepted by $\nbw_\g$, which holds iff $\KrFamily\models \f$.

For a model $\Kr$, we denote by $\Kr \cap_k \nbw_\g$ the intersection of $\Kr$ and $\nbw_\g$ w.r.t. {$\AP\times \{\pi_k\}$}, 
taking the projection over $\AP\times \{\pi_i\}_{i=1}^{k-1}$.
Thus, $\lang(\Kr \cap_k \nbw_\g)$ 
is the set of all $(k-1)$-words that 
{\em an extension} (i.e., $\exists$) by a word in $\lang(\Kr)$ to a $k$-word in $\lang(\nbw_\g)$.
Oppositely, 
{$\lang(\overline{\Kr\cap_k \overline{\nbw_\psi}})$} 
is the set of all $(k-1)$-words that 
{\em every extension} (i.e., $\forall$) by a $k$-word in $\lang(\Kr)$ 
is in {$\lang({\nbw_\g})$}. 
%


We first construct NBWs {$\nbw_2,\ldots, \nbw_{k-1}, \nbw_{k}$}, 
%
such that for every {$1< i < k$}, we have 
{$\tupleof{K_1,\ldots, K_i}\models (\alpha_i, A_{i+1})$} 
iff 
$\KrFamily\models (\alpha,A_\g)$, where $\alpha_i = \quant_1\pi_1\dots \quant_i\pi_i$.

For $i = k$, if $\quant_k = \exists$, then $A_k = \Kr_k\cap_k A_\g$; 
otherwise  if $\quant_k = \forall$, {$\nbw_k = \overline{\Kr_k\cap_k \overline{A_{\g}}}$.}
%
For $1< i< k$, 
if $\quant_i = \exists$ then $\nbw_i = \Kr_i\cap_i A_{i+1}$; 
otherwise if $\quant_i = \forall$,  
$\nbw_i = \overline{\Kr_i\cap_i{\overline{A_{i+1}}}}$. 
Then, for every {$1< i < k$}, 
we have $\tupleof{\Kr_1,\ldots, \Kr_i}\models (\alpha_i, \nbw_{i+1})$ 
iff $\tupleof{\Kr_1,\ldots, \Kr_k}\models \f$.

We now prove by induction on $k$ that $\KrFamily\models \f$ iff $\tupleof{C_1,\ldots C_k}\models \f$.
For $k=1$, it holds that $\KrFamily\models \f$ 
iff 
$\Kr_1\models (\quant_1\pi_1,\nbw_2)$. 
If $\quant_1 = \forall$, then $\Kr_1\models (\quant_1\pi_1,\nbw_2)$ iff $\Kr_1\cap \overline{\nbw_2}= \emptyset$.
If $\quant_1 = \exists$, then $\Kr_1\models (\quant_1\pi_1,\nbw_2)$ iff $\Kr_1\cap \nbw_2\neq \emptyset$.
In both cases, a lasso witness to the non-emptiness exists.
For $1<i<k$, we prove that $\tupleof{C_1, \ldots, C_i, \Kr_{i+1}}\models (\alpha_{i+1}, A_{i+2})$ iff 
$\tupleof{C_1, \ldots, C_i, C_{i+1}}\models (\alpha_{i+1}, A_{i+2})$.
If $\quant_i = \forall$, then the first direction simply holds because  
{$\lang(C_{i+1})\subseteq \\ \lang(\Kr_{i+1})$}.
For the second direction, every extension of $c_1,c_2,\ldots c_{i}$ 
(i.e., lassos in $C_1,C_2,\ldots C_i$) by a path $\tau$ in $\Kr_{i+1}$ is in {$\lang(\nbw_{i+2})$}.
%
Indeed, otherwise we can extract a lasso $c_{i+1}$ 
such that $c_1,c_2,\ldots c_{i+1}$ is in {$\overline{\lang(A_{i+2})}$}, a contradiction. 
%
If $\quant_i = \exists$, then $\lang(C_{i+1})\subseteq \lang(\Kr_{i+1})$ implies the second direction.
For the first direction,
we can extract a lasso $c_{i+1}\in \lang(C_{i+1})$ such that 
{$\tupleof{c_1,c_2,\ldots c_i,c_{i+1}}\in \lang(\nbw_{i+2})$}.
\qed
\end{proof} 

One can use Theorem~\ref{thm:lassos} and the observations above 
to construct a sound and complete BMC algorithm for 
{both $\forall\exists$ and $\exists\forall$ hyperproperties.}
%
Indeed, consider a multi-model $\tupleof{\Kr_1,\Kr_2}$, 
and a hyperproperty $\f = \forall \pi. \exists \pi'. ~\g$. 
Such a BMC algorithm would try and verify $\tupleof{\Kr_1,\Kr_2}\models\f$ directly, or try and prove  $\tupleof{\Kr_1,\Kr_2}\models\neg\f$. 
In both cases, a run {\tacas may find} a short lasso example for the model under $\exists$ 
($\Kr_2$ in the former case and $\Kr_1$ in the latter), leading to a shorter run.
%
However, in both cases, the model under $\forall$  would have to be explored to the maximal lasso length implicated by Theorem~\ref{thm:lassos}, which is doubly-exponential. Therefore, this naive approach would be highly inefficient.


%% file: algos.tex
\section{Simulation-Based BMC Algorithms for HyperLTL}
\label{sec:simulation}

We now introduce efficient simulation-based BMC algorithms for
verifying hyperproperties of the types 
$\forall \pi. \exists \pi'.\Box \pred$ and
$\exists \pi. \forall \pi'.\Box \pred$, where $\pred$ is a {\em relational predicate}
(a predicate over a pair of states).
The key observation is that simulation naturally induces the
exploration of infinite traces without the need to explicitly unroll
the structures, and without needing to synchronize the indices of the
symbolic variables in both traces.
Moreover, in some cases our algorithms allow to only partially explore
the state space of a Kripke structure and give a conclusive answer {\tacas efficiently}.

Let 
{$\Mp = \tupleof{S_P, S_{P}^\init, \trans_P,$ $\AP_P, L_P} $ and 
$\Mq = \tupleof{S_Q, S_{Q}^\init, \trans_Q,$ $\AP_Q, L_Q} $}
be two
Kripke structures, and consider a hyperproperty of the form
$\forall \pi. \exists \pi'.~\Box \pred$.
Suppose that there exists a simulation from $\Mp$ to $\Mq$. Then,
every trace in $\Mp$ is embodied in $\Mq$. 
Indeed, we can show by
induction that for every trace 
$\ptrace = \pstate{1}\pstate{2}\ldots$ in $\Mp$, 
there exists a trace $\qtrace = \qstate{1}\qstate{2}\ldots$ in $\Mq$, 
such that $\qstate{i}$ simulates $\pstate{i}$ for every $i\geq 1$; 
therefore, $\ptrace$ and $\qtrace$ are equally labeled.  
We generalize the labeling constraint in the
definition of standard simulation by requiring, given $\pred$, that if
$(s_p, s_q)$ is in the simulation relation, then $(s_p,s_q)\models\pred$. We
denote this generalized simulation by $\aesim$.
Following similar considerations, we now have that for every trace
$\ptrace$ in $\Mp$, there exists a trace $\qtrace$ in $\Mq$ such that
$(\ptrace,\qtrace)\models\Box\pred$.
Therefore, the following result holds:

\begin{lemma}
  \label{lem:AE}
  Let $\Mp$ and $\Mq$ be Kripke structures, and let
  $\f =\forall\pi.\exists\pi'.~\Box \pred$ be a \HyperLTL formula. If there exists $\aesim$ from $\Mp$ to $\Mq$, then $\tupleof{\Mp,\Mq}\models \f$.
\end{lemma}

We now turn to properties of the type $\exists\pi. \forall \pi'.~\Box \pred$. In this case, we must find a single trace in $\Mp$ that matches {\tacas every trace in $\Mq$.} 
Notice that $\aesim$ (in the other direction) does not suffice, since it is not guaranteed that the same trace in $\Mp$ is used to match all traces in $\Mq$. 
However, according to Theorem~\ref{thm:lassos}, it is guaranteed that if $\tupleof{\Mp,\Mq}\models \exists \pi. \forall \pi'.~\Box \pred$, then there exists such a single {lasso} trace $\ptrace$ in $\Mp$ {\tacas as the witness of the satisfaction}. 
We therefore define a second notion of simulation, denoted $\easim$, as follows. 
Let $\ptrace = \pstate{1} \pstate{2} \ldots \pstate{n} \ldots \pstate{\ell}$ be a lasso trace in $\Mp$ 
(where $\pstate{\ell}$ closes to $\pstate{n}$, that is, $(\pstate{\ell}, \pstate{n}) \in  \trans_P$). 
A relation $\simrelation$ from $\ptrace$ to $\Kr_Q$ is considered as a $\easim$ from $\ptrace$ to $K_Q$, 
if the following holds: 
\begin{enumerate}
    \item $(s_p,s_q)\models\pred$ for every $(s_p,s_q)\in\simrelation$.
    \item $(\pstate{1},s_q)\in\simrelation$ for every $s_q \in S_{Q}^\init$.
    \item If $(\pstate{i},s_q(i))\in \simrelation$, then for every successor $s_q(i+1)$ of $s_q(i)$, it holds that \\ $(\pstate{i+1},s_q(i+1))\in\simrelation$ (where $\pstate{\ell+1}$ is defined to be $\pstate{n}$).
\end{enumerate}
If there exists a lasso trace $\ptrace$, then we say that there exists $\easim$ from $\Mp$ to $\Mq$.
Notice that the third requirement in fact unrolls $\Mq$ in a way that guarantees that for every trace $\qtrace$ in $\Mq$, it holds that $(\ptrace,\qtrace) \models \Box\pred$.
Therefore, the following result holds:



\begin{lemma}
  \label{lem:EA}
  Let $\Mp$ and $\Mq$ be Kripke structures, and let
  $\f = \exists\pi.\forall\pi'.~\Box \pred$. If there exists a $\easim$ from $\Mp$ to $\Mq$, then $\tupleof{\Mp,\Mq}\models \f$.
\end{lemma}


Lemmas~\ref{lem:AE} and~\ref{lem:EA} enable sound algorithms for
model-checking 
$\forall \pi. \exists \pi'.~\Box\pred$ and
$\exists \pi. \forall \pi'.~\Box\pred$ hyperproperties with loop conditions.
To check the former, check whether there exists $\aesim$ from $\Mp$
to $\Mq$; to check the latter, check for a lasso trace $\ptrace$ in
$\Mp$ and $\easim$ from $\ptrace$ to $\Mq$.
Based on these ideas, we introduce now two SAT-based BMC algorithms.

For $\forall\exists$ hyperproperties, we not only check for the
existence of $\aesim$, but also iteratively seek a small subset of $S_Q$ that
suffices to simulate all states of $S_P$. 
While finding $\aesim$, as for standard simulation, is polynomial, the problem of finding a
simulation that uses a bounded number of $K_Q$ states is NP-complete {\tacas (see Appendix~\ref{sec:bounded.sim} for details).}
%
This allows us to
efficiently handle instances in which $\Mq$ is large.
Moreover, {\tacas we introduce in Subsection~\ref{subsec:prophecies}}
the use of {\em prophecy variables}, allowing us to overcome cases in
which the models satisfy the property but $\aesim$ does not exist.

For $\exists\forall$ hyperproperties, we search for $\easim$ by
seeking a lasso trace $\ptrace$ in $\Mp$, whose length increases with
every iteration, similarly to standard BMC techniques for \LTL. 
Of course, in our case, $\ptrace$ must be matched with
the states of $\Mq$ in a way that ensures $\easim$. In the worst case,
the length of $\ptrace$ may be doubly-exponential in the sizes
of the systems. However, as our experimental results show, in case of
satisfaction the process can terminate much sooner.

We now describe our BMC algorithms and our SAT encodings in detail.
First, we fix the unrolling depth of $\Mp$ to $n$ and of $\Mq$ to
$k$.
%
%
%
To encode states of $K_P$ we allocate a family of Boolean variables
$\{x_i\}_{i=1}^n$.
Similarly, we allocate $\{y_j\}_{j=1}^k$ to represent the states
of $K_Q$.
Additionally, we encode the simulation relation $\simulate$ by creating
$n\times{}k$ Boolean variables $\{\SIM_{ij}\}_{i=1}^n,_{j=1}^k$ such
that $\SIM_{ij}$ holds if and only if $\simulate(p_i,q_j)$.
We now present the three {\tacas variations} of encoding: 
(1) EA-Simulation ($\easim$), 
(2) AE-Simulation ($\aesim$), and 
(3) a special variation where we enrich AE-Simulation with prophecies.

\subsection{Encodings for EA-Simulation}
\label{subsec:EA}
The goal of this encoding is to find a lasso path $\ptrace$ in $\Mp$
that guarantees that there exists $\easim$ to $\Mq$.
Note that the set of states that $\ptrace$ uses may be much smaller than the whole
of $K_P$, while the state space of $K_Q$ must be explored exhaustively.
%
%
We force $x_0$ be an initial state of $K_P$ 
and for $x_{i+1}$ to follow
$x_i$ for { every $i$} we use, but for $K_Q$ we will let the solver fill
freely each $y_k$ and add constraints\footnote{An alternative is to
fix an enumeration of the states of $K_Q$ and force the assignment of
$y_0\ldots$ according to this enumeration instead of constraining a
symbolic encoding, but the explanation of the symbolic algorithm
above is simpler.} for the full exploration of $K_Q$.

\begin{itemize}
\item {\bf All states are legal states.}  The solver must only search
  legal encodings of states of $K_P$ and $K_Q$ (we use $\Mp(x_i)$ to
  represent the combinations of values that represent a legal state in
  $S_P$ and similarly $\Mq(y_j)$ for $S_Q$):
\begin{align}
	\bigwedge\limits_{i=1}^n 
	\Mp(x_i) 
	\land 
	\bigwedge\limits_{j=1}^k 
	\Mq(y_j)
  \label{EAone}
\end{align}

\item {\bf  Exhaustive exploration of $K_Q$.}
  {\tacas
  We require that two different indices $y_j$ and $y_r$ represent two different
  states in $K_Q$, so if $k=|K_Q|$, then all states are represented 
  {(note that the validity of states is implied by (1))}:
 }
{\tacas
\begin{align}
	\bigwedge\limits_{j\neq{}r}
	(\M_\BBB(y_j) \land \M_\BBB(y_r)) 
	\rightarrow
	(y_j \neq y_r)
	\label{EAtwo}
\end{align}}%
{\tacas{}%
\noindent where $y_j\neq y_r$ captures that some bit
distinguishes the states encoded by $j$ and $r$.}

\item {\bf The initial $S_{P}^\init$ state simulates all initial $S_{Q}^\init$ states.}
  %
 {\tacas
 	State $x_0$ is an initial state of $K_P$ and simulates all initial states of $K_Q$ 
 	(we use $I_P$ to represent a legal initial state in $K_P$ and $I_Q(y_j)$ 
 	for $S_{Q}^\init$ of $K_Q$):
 }	
  \begin{align}
      I_P(x_0) \And
    \big(\bigwedge\limits_{j=1}^k
    I_\BBB(y_j) 
    \Into 
    \simulate(x_0, y_j)
    \big)
    \label{EAthree}
\end{align}


\item {\bf Successors in $K_Q$ are simulated  by successors in $K_P$.}
  We first introduce the following formula {\tacas $\SUCC_T(x,x')$} to capture
  one-step of the simulation, that is, $x'$ follows $x$ and for all
  $y$ if $T(x,y)$ then $x'$ simulates all successors of $y$ 
  {\tacas (we use $\delta_Q(y, y')$ to  represent that $y$ and $y'$ states are in $\delta_{Q}$ of $\Kr_Q$, similarly for $(x, x') \in \delta_{P}$ of $K_P$ we use $\delta_{P}(x, x'))$ }:
  \[
    \SUCC_T(x,x')\DefinedAs
    \bigwedge\limits_{y=y_1}^{y_k} T(x, y)
    \rightarrow \big(
    \bigwedge\limits_{y'=y_1}^{y_k} \delta_{\BBB} (y,y')
    \rightarrow T(x', y')
    \big)
  \]
  We can then define that $x_{i+1}$ follows $x_i$:
\begin{align}
  \bigwedge\limits_{i=1}^{n-1}\big[
	\delta_{\AAA} (x_i, x_{i+1}) \And
  \SUCC_T(x_i,x_{i+1})\big]
    \label{EAfour}
\end{align}
And, $x_n$ has a jump-back to a  previously seen state:
\begin{align}
  \bigvee\limits_{i=1}^n
  \big[
	\transit{P} (x_n, x_i) \And
  \SUCC_T(x_n,x_i)
  \big]
  \label{EAfive}
\end{align}

\item {\bf Relational state predicates are fulfilled by simulation.}
  Everything relating in the simulation fits the relational predicate,
  defined as a function $\pred$ of two sets of labels
  {\tacas (we use $L_Q(y)$ to represent the set of labels on the $y$-encoded state in $K_Q$, similarly, $L_P(x)$ for the $x$-encoded state in $K_P$)}:
\begin{align}
	\bigwedge\limits_{i=1}^n 
	\bigwedge\limits_{j=1}^k 
	\simulate(x_i,y_j) 
	\rightarrow  
	\pred(L_\AAA(x_i), L_\BBB(y_j))
	\label{EAsix}
\end{align}
\end{itemize}
We use $\varphiEAnk$ for the SAT formula that results of conjoining
$(\ref{EAone})$-$(\ref{EAsix})$ for bounds $n$ and $k$.
If $\varphiEAnk$ is satisfiable, then there exists $\easim$ from $\Mp$ to $\Mq$.

\commentout{
The following lemma guarantees the soundness of the technique.

\begin{proof}(sketch) Assume $\varphiEAnk$ is satisfiable and let
  $\ptrace$ be the path that corresponds to the assignment given by the
  SAT solver to $(x_1\ldots x_n)(x_j\ldots x_n)^\omega$, where $j$ is
  obtained by $(\ref{EA5})$.
  The proof proceeds by induction showing that for all naturals $i$,
  the first $i$ steps of $\ptrace$ match the first $i$ steps of every
  path of $\Mq$, according to $\pred$.
  \qed
\end{proof}
}

\subsection{Encodings for AE-Simulation}
\label{subsec:AE}
Our goal now is to find a set of states $S_\BBB' \subseteq S_\BBB$ that is
able to simulate all states in $K_\AAA$.
Therefore, as in the previous case, the state space $K_\AAA$
corresponding to the $\forall$ quantifier will be explored
exhaustively, and so $n=|K_\AAA|$, while $k$ is the number of states in
$K_\BBB$, which increases in every iteration. As we have explained, this
allows finding a small subset of states in $K_\BBB$ which suffices to simulate all
states of $\Mp$.

%
%
%

\begin{itemize}
\item {\bf All states in the simulation are legal states.}  Again,
  every state guessed in the simulation is a legal state from $K_\AAA$
  or $K_\BBB$:
\begin{align}
	\bigwedge\limits_{i=1}^n 
	\M_\AAA(x_i) 
	\land 
	\bigwedge\limits_{j=1}^k 
  	\M_\BBB(y_j)
  \tag{$1'$}
  \label{AEone}
\end{align}

\item {\bf $K_\AAA$ is exhaustively explored.}  Every two different
  indices in the states of $K_\AAA$ are different states\footnote{As in the previous
    case, we could fix an enumeration of the states of $S_P$ and fix
    $x_0x_1\ldots$ to be the states according to the enumerations.}:
{\tacas
\begin{align}
	\bigwedge_{i \neq r}  ( \M_P(x_i) \land \M_P(x_r)) \rightarrow (x_i \neq x_r) 
	\tag{$2'$}
	\label{AE2}
\end{align}
}

\item {\bf All initial states in $K_\AAA$ must match with some initial state in $K_\BBB$.}
  Note that, contrary to the $\exists\forall$ case, 
  here the initial state
  in $K_Q$ may be different for each initial state in $S_P$:
\begin{align}
	\bigwedge\limits_{i=1}^n
	\bigvee\limits_{j=1}^k 
	I_\AAA(x_i) 
	\rightarrow 
	\big(
	I_\BBB (y_j) 
	\land
	\simulate(x_i, y_j) 
  \big)
  \tag{$3'$}
  \label{AE3}
\end{align} 
	
\item {\bf For every pair in the simulation, each successor in $K_\AAA$
    must match with some successor in $K_\BBB$.}
 For each $(x_i, y_j)$ in the simulation, every successor state of
  $x_i$ has a matching successor state of $y_j$:
\begin{align}
	\bigwedge\limits_{i=1}^n
	\bigwedge\limits_{t=1}^n 
	\transit{\AAA}(x_i, x_t)
	\rightarrow 
	\bigwedge\limits_{j=1}^k 
	\Big[
	\simulate(x_i, y_j) 
	\rightarrow 
	\bigvee\limits_{r=1}^k  
	\Big(	
	\transit{\BBB}(y_j, y_r)
	\land 
	\simulate(x_t, y_r)
	\Big)
  \Big]
  \tag{$4'$}
  \label{AE4}
\end{align}	

\item {\bf Relational state predicates are fulfilled.}
Similarly, all pairs of states in the simulation should respect the relational $\pred$:
 {\tacas
\begin{align}
	\bigwedge\limits_{i=1}^n 
	\bigwedge\limits_{j=1}^k 
	\simulate(x_i,y_j) 
	\rightarrow  
  \pred(L_\AAA(x_i), L_\BBB(y_j))
  \tag{$5'$}
  \label{AEfive}
\end{align}
}

\end{itemize}
We now use $\varphiAEnk$ for the SAT formula that results of
conjoining (\ref{AEone})-(\ref{AEfive}) for bounds $n$ and $k$.
If $\varphiAEnk$ is satisfiable, then there exists $\aesim$ from $\Mp$ to $\Mq$.
%
%

\input{proph}


%% file: proph.tex
\subsection{Encodings for AE-Simulation with Prophecies}
\label{subsec:prophecies}

{\tacas 
	The AE-simulation encoding introduced in Section~\ref{subsec:AE} can handle most properties of the form
	 $\forall\exists \G \pred$; 
	however, it is unable to cope well with systems 
	(in particular the system $K_P$ for the $\forall$ quantifier) that exhibit non-determinism.}
The reason, as illustrated in the following example, is that the
simulation matches immediately the successor for the $\exists$ path
without inspecting the future of the $\forall$ path.
This is illustrated in the following example.
\begin{example}
  Consider Kripke structures $K_1$ and $K_2$ from
  Section~\ref{sec:intro}, and \hltl formula {\tacas
    $\varphi_2 = \forall \pi. \exists \pi'.~ \G (a_{\pi}
    \leftrightarrow a_{\pi'})$.  }
  It is easy to see that the two models satisfy {\tacas $\varphi_2$}, since
  mapping the sequence of states $(s_1s_2s_3)$ to $(q_1q_2q_4)$ and
  $(s_1s_2s_4)$ to $(q_1q_3q_5)$ guarantees that the matched paths
  satisfy {\tacas $\Always(a_{\pi} \Iff a_{\pi'})$}.
  However, the technique in Section~\ref{subsec:AE} cannot
  differentiate the occurrences of $s_2$ in the two different cases.
  \qed
\end{example}

To solve this, we incorporate the notion of {\em prophecies} to our setting.
Prophecies have been proposed as a method to aid in the verification
of hyperliveness~\cite{coenen19verifying}
(see~\cite{beutner22prophecy} for a systematic method to construct
these kind of prophecies).
For simplicity, we restrict here to prophecies expressed as safety automata.
A safety prophecy over $\AP$ is a Kripke structure $U=\ktuple$, 
such that $\Traces(U)=\AP^\omega$. 
The product $K\times U$ of a Kripke structure $K$ with a prophecy $U$
preserves the language of $K$ (since the language of $U$ is
universal).
Recall that in the construction of the product, states
{\tacas $(s,u)\in (\M \times U)$} 
that have incompatible labels are removed.
The direct product can be easily processed by repeatedly removing dead states, resulting in a Kripke
structure $K'$ whose language is $\Traces(K')=\Traces(K)$.
Note that there may be multiple states in $K'$ that correspond to
different states in $K$ for different prophecies.
The prophecy-enriched Kripke structure can be directly passed to the
method in Section~\ref{subsec:AE}, so the solver can search for a
\aesim that takes the value of the prophecy into account.

\begin{figure}[t]
	\begin{tabular}{l@{\hspace{4em}}l}
		\hspace{4.5em}  \includegraphics[scale=0.4]{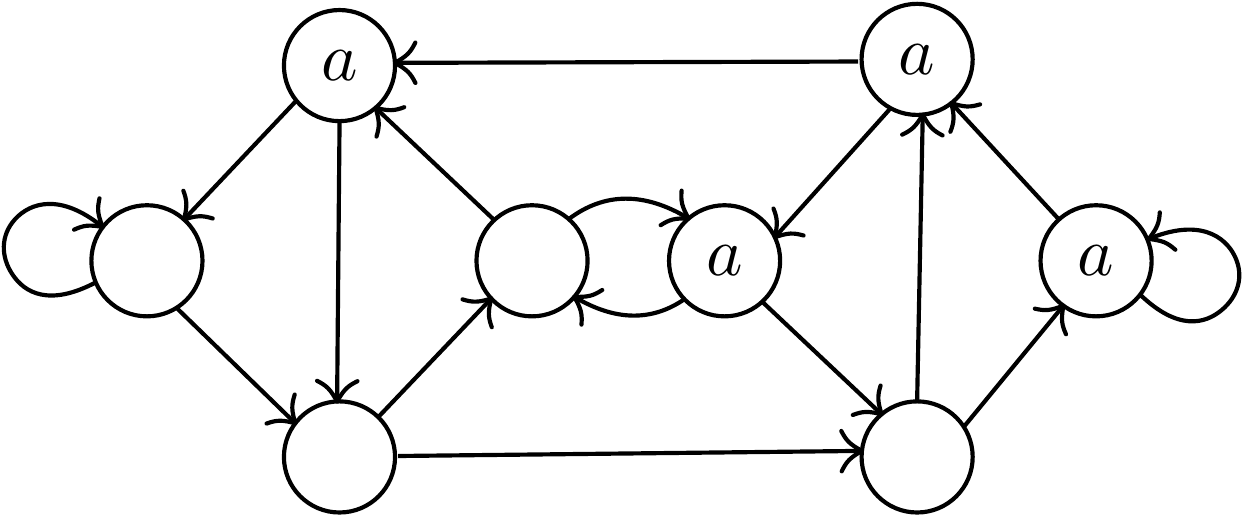} &
		\includegraphics[scale=0.4]{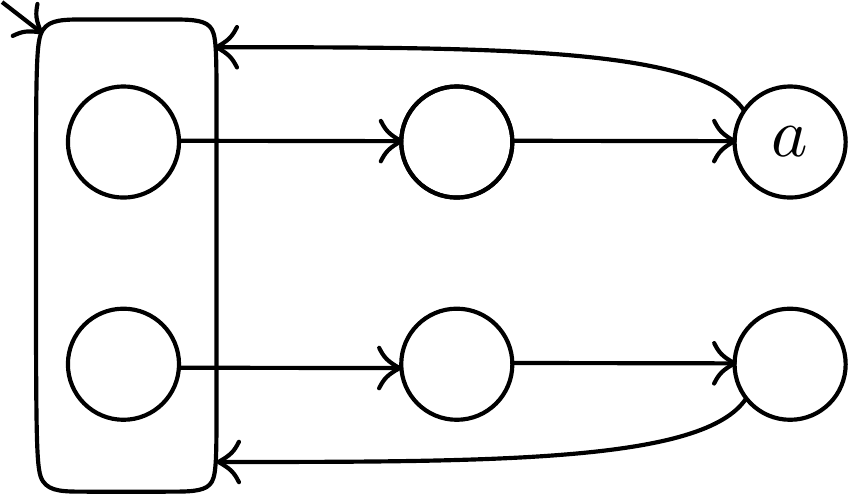}
	\end{tabular}                                                  
	\caption{Prophecy automaton for $\Next\Next a$ (left) and
		its composition with $K_1$ (right).}
	\label{fig:proph}
\end{figure}
\begin{example}
  Consider the {\tacas prophecy automaton shown in Fig.~\ref{fig:proph} (left)},
  where all states are initial. 
%
  Note that for every state, either all its successors are labeled
  with $a$ (or none are), and all successors of its successors are
  labeled with $a$ (or none are).
In other words, this structure encodes the prophecy $\Next\Next a$.
The product $K_1'$ of $K_1$ with the prophecy automaton $U$ for
$\Next\Next a$ is shown in Fig.~\ref{fig:proph} (right).
Our method can now show that $\tupleof{K_1',K_2}\models\varphi_2$, since it
can distinguish the two copies of $s_1$ (one satisfies $\Next\Next a$
and is mapped to $(q_1q_2q_4)$, while the other is mapped to $(q_1q_3q_5)$).
\qed

\end{example}


%% file: exp.tex
\section{Implementation and Experiments}
\label{sec:exp}

\newcommand{\violation}{\text{violation}}

We have implemented our algorithms using the SAT solver \code{Z3} through its python
API \code{Z3Py}~\cite{moura2008z3}.
The SAT formulas introduced in Section~\ref{sec:simulation} are encoded into the two scripts \texttt{simEA.py} and \texttt{simAE.py}, for finding simulation relations for the $\easim$ and $\aesim$ cases, respectively. 
We evaluate our algorithms with a set of experiments, which includes all forms of quantifiers
with different sizes of given models, {\tacas as presented earlier in Table~\ref{tab:category}}.
%
Our simulation algorithms benefit the most in the cases of the form
$\forall_\msmall~\exists_\mbig$. 
When the second model is
substantially larger than the first model, $\aesim$ is able to
prove that a $\forall\exists$ hyperproperty holds by exploring only a
subset of the second model.
In this section, besides $\forall_\msmall~\exists_\mbig$ cases, we also investigate multiple cases on each category in Table~\ref{tab:category} to demonstrate the generality and applicability of our algorithms.   
All case studies are run on a MacBook Pro with Apple M1 Max chip and 64 GB of memory.

\subsection{{\tacas Case Studies and Empirical Evaluation}}

\subsubsection{Conformance in Scenario-based Programming.}
%
In scenario-based programming, scenarios provide a big picture of the
desired behaviors of a program, and are often used in the context of
program synthesis or code generation.
A synthesized program should obey what is specified 
in the given set of scenarios to be considered {\em correct}.
That is, the program {\em conforms} with the scenarios.
%
The conformance check between the scenarios and the 
synthesized program can be specified as a $\forall\exists$-hyperproperty: 
$$ 
\varphi_\textsf{conf} = \forall \pi. \exists \pi'. \bigwedge_{p \in \AP}\G\ ({p}_\pi \leftrightarrow 
{p}_{\pi'}),
$$
where $\pi$ is over the scenario model and $\pi'$ is over the synthesized program. 
That is, for all possible runs in the scenarios, there must exists a run in the program, such that their behaviors always match.
%

We look into the case of synthesizing an {\em Alternating Bit Protocol (ABP)} from four given scenarios, inspired by~\cite{alur2014synthesizing}.
ABP is a networking protocol that guarantees reliable message transition, when message loss or data duplication are possible.
The protocol has two parties: \sender\ and \receiver, which can take three different actions: \send, \receive, and \wait. Each action also specifies which message is currently transmitted: either a {\em packet} or {\em acknowledgment}
{\tacas (see~\cite{alur2014synthesizing} for more details)}. 
%
The correctly synthesized protocol should not only have complete functionality but also {\em include all scenarios}. That is, for every trace that appears in some scenario, there must exist a corresponding trace in the synthesized protocol. 
By finding $\aesim$ between the scenarios and the synthesized protocols, we can prove the conformance specified with $\varphi_\textsf{conf}$.
Note that the scenarios are often much smaller than the actual synthesized protocol, and so 
this case falls in the $\forall_\msmall~\exists_\mbig$ category in Table~\ref{tab:category}.
%
We consider two variations: a correct and an incorrect ABP (that cannot handle packet loss).
Our algorithm successfully identifies 
 a $\aesim$ that satisfies $\varphi_\textsf{conf}$ for 
the correct ABP, and returns UNSAT for the incorrect protocol, since 
the packet loss scenario cannot be simulated. 
%




\subsubsection{Verification of Model Translation.}
It is often the case that in model translation (e.g., compilation),
solely reasoning about the source program does not provide guarantees about the desirable behaviors in the target executable code. 
Since program verification is expensive
compared with repeatedly checking the target,
alternative approaches such as {\em certificate translation}~\cite{barthe2009implementing} are often preferred.
Certificate translation takes inputs of a high-level program (source) with a given specification, and computes a set of verification conditions (certificates) for the low-level executable code (target) to prove that a model translation is safe. 
However, this technique still requires extra efforts to map the
certificates to a target language, and the size of generated
certificates might explode quickly (see~\cite{barthe2009implementing} for retails).
We show that our simulation algorithm can directly show the correctness 
of a model translation more efficiently by investigating the 
source and target with the same formula $\varphi_\textsf{conf}$ used for ABP.
%
%
That is, the specifications from the source runs $\pi$ are always preserved in some target runs $\pi'$, which infers a correct model translation.
Since translating a model into executable code implies adding extra
instructions such as writing to registers, it also falls into the 
$\forall_\msmall~\exists_\mbig$ category in Table~\ref{tab:category}.

We investigate a program from~\cite{barthe2009implementing} that performs 
{\em matrix multiplication (MM)}.
%
%
When executed, the C program is translated from high-level code (C)  to low-level code RTL (Register Transfer Level), which contains extra steps to read from/write to memories. 
Specifications are triples of $\langle \Pre, \annot, \Post \rangle$,
where $\Pre$, and $\Post$ are assertions and $\annot$ is a partial
function from labels to assertions (see~\cite{barthe2009implementing}
for detailed explanations).
The goal is to make sure that the translation does not violate the
original verified specification.
%
In our framework, instead of translating the certification, we find a simulation that 
satisfies {\tacas $\varphi_\textsf{conf}$}, proving that the translated code also satisfies the specification. 
We also investigate two variations in this case: a correct translation and an incorrect 
translation, and our algorithm returns SAT (i.e., finds a correct \aesim simulation) in the former case, and returns UNSAT for the latter case.

\begin{wrapfigure}{r}{.4 \columnwidth}
	\vspace*{-7mm}
	\centering
	\begin{minipage}{.4 \columnwidth}
			\includegraphics[scale=.16]{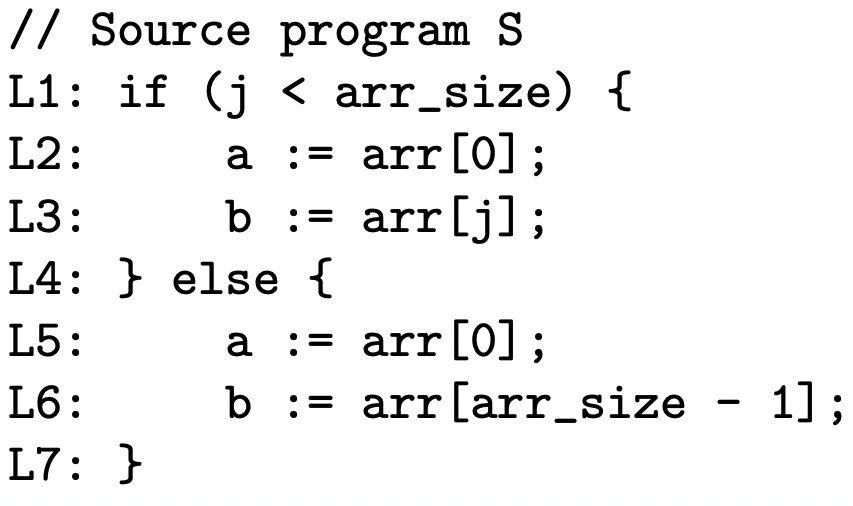}
	\end{minipage}
	\begin{minipage}{.4 \columnwidth}
		\includegraphics[scale=.16]{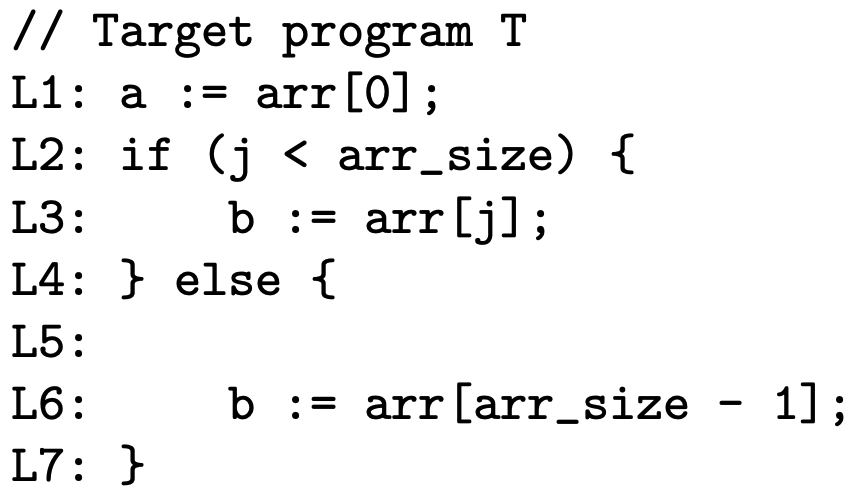}
	\end{minipage}
	\caption{The common branch factorization example~\cite{namjoshi2020witnessing}.}
	\label{fig:cbf}
	\vspace*{-5mm}
\end{wrapfigure}
\subsubsection{Compiler Optimization.}
Secure compiler optimization aims at preserving input-output behaviors of an original implementation and a target program after applying optimization techniques, including security policies. 
The conformance between source and target programs guarantees that the optimizing procedure does not introduce vulnerabilities such as information leakage.
Furthermore, optimization is often not uniform for the same source, because one might compile the source to multiple different targets with different optimization techniques.
As a result, an efficient way to check the behavioral equivalence between the source and target 
provides a correctness guarantee for the compiler optimization.

Imposing optimization usually results in a smaller program. 
For instance, {\em common branch factorization} (CBF) finds common operations in an if-then-else 
structure, and moves them outside of the conditional so that such operation is only executed once.
As a result, for these optimization techniques, checking the conformance 
of the source and target falls in the $\forall_\mbig~\exists_\msmall$  category.
That is, given two programs, source (\mbig) and target (\msmall), we check the following formula:
$$ \varphi_\textsf{sc} = \forall \pi. \exists \pi'. ~ (\textsf{in}_\pi \leftrightarrow \textsf{in}_{\pi'}) \rightarrow \G\ (\textsf{out}_\pi \leftrightarrow \textsf{out}_{\pi'}).$$


In this case study we investigate the strategy {\tacas CBF} using the example in Figure~\ref{fig:cbf} inspired by~\cite{namjoshi2020witnessing}.
%
We consider two kinds of optimized programs for the strategy, one is the correct 
optimization, one containing bugs that violates the original behavior due to the optimization.
For the correct version, our algorithm successfully discovered a simulation relation 
between the source and target, and the simulation relation 
returns a smaller subset of states in the second model (i.e., $|Q'| < |Q|$).
For the incorrect version, we received UNSAT.
%


\begin{wrapfigure}{r}{0.18 \columnwidth}
	\vspace*{-7mm}
	\scalebox{.6}{
		\input{fig_rob.tex}
	}
	\caption{A\\ robust paths.}
	\label{fig:robot}
	 \vspace*{-8mm}
\end{wrapfigure}
\subsubsection{Robust Path Planning.}
In robotic planning, {\em robustness planning (RP)} refers to a path that is able to consistently complete a mission without being interfered by the uncertainty in the environment (e.g., adversaries). 
For instance, in the 2-D plane in Fig.~\ref{fig:robot}, an agent is trying to go from the 
starting point (blue grid) to the goal position (green grid). 
%
The plane also contains three adversaries on the three corners other
than the starting point (red-framed grids), and {\tacas the
  adversaries move trying to catch the agent but can only move in one
  direction (e.g., clockwise).}
This is a $\exists_\msmall~ \forall_\mbig$ setting, since the adversaries may have 
several ways to cooperate and attempt to catch the agent.
We formulate this planning problem as follows:
$$ \varphi_\textsf{rp} = \exists \pi. \forall \pi'.~ \G\ (\textsf{pos}_\pi \not\leftrightarrow \textsf{pos}_{\pi'}).$$
That is, there exists a robust path for the agent to safely reach the goal regardless of all the ways that the adversaries could move.
We consider two scenarios, one in which there exists a way for the agent to form a robust path and one does not. 
Our algorithm successfully returns SAT for case which the agent can form a robust path, 
and returns UNSAT for which a robust path is impossible to find.

\subsubsection{Plan Synthesis.}
The goal of {\em plan synthesis (PS)} is to synthesize a single comprehensive plan that can simultaneously
satisfy all given small requirements has wide application in planning
problems.
We take the well-known toy example, {\em wolf, goat, and
  cabbage}\footnote{\url{https://en.wikipedia.org/wiki/Wolf,_goat_and_cabbage_problem}},
as a representative case here.
The problem is as follows. A farmer needs to cross a river by boat with a wolf, a goat, and a cabbage.
However, the farmer can only bring one item with him onto the boat each time. 
In addition, the wolf would eat the goat, and the goat would eat the
cabbage, if they are left unattended. 
The goal is to find a plan that
allows the farmer to successfully cross the river with all three
items safely.
A plan requires the farmer to go back and forth with the boat
with certain possible ways to carry different items, while 
all small requirements (i.e., the constraints among each item) always satisfied.
In this example, the overall plan is a big model while the requirements form  
a much smaller automaton.
Hence, it is a $\exists_\mbig~ \forall_\msmall$ problem that can be specified with 
the following formula:
$$ \varphi_\textsf{ps} = \exists \pi. \forall \pi'.~\G\ ( \textsf{action}_\pi \not\leftrightarrow \textsf{violation}_{\pi'}).$$
%

\begin{table}[t]
	\centering
	\input{table_exp}
	\caption{
		{\tacas Summary of our case studies. 
			The outcomes with simulation discovered 
			show how our algorithms find a smaller subset for either $K_P$ or $K_Q$.
		}
        }
 	\label{tab:summary}
 	\vspace*{-8mm}
\end{table}
\subsection{Analysis and Discussion}
The summary of our empirical evaluation is presented in Table~\ref{tab:summary}.
%
For the $\forall\exists$ cases, our algorithm successfully
{\tacas finds a set $|S_Q'| < |S_Q|$ that satisfies the properties for the cases ABP and CBF.}
Note that case MM does not find a small subset, since we manually add
extra {\em paddings} on the first model to align the length of both
traces. 
We note that handling this instance without padding requires asynchornicity---
a much more difficult problem, which we leave for future work. 
%
For the $\exists\forall$ cases, we are able to find a subset of $S_P$
which forms a single lasso path that can simulate all runs in $S_Q$ for 
all cases RP and GCW.
{\tacas We emphasize here that previous BMC techniques 
	(i.e., $\HyperQube$) cannot handle most of the cases in Table~\ref{tab:summary}
	due to the lack of loop conditions.}

%% file: fig_rob.tex
\begin{tikzpicture} 
    [
        box/.style={rectangle,draw=gray,thick, minimum size=1cm}, 
    ]

\foreach \x in {0,1,...,2}{
    \foreach \y in {0,1,...,2}
        \node[box] at (\x,\y){};
}

\node[box,fill = blue!60] at (0,0) {};

\node[box,fill = green] at (2,2) {};
  

 \draw[arw, ->, line width=0.5mm, color = blue](0,0)--(1,0);
 \draw[arw, ->, line width=0.5mm, color = blue](1,0)--(1,1);

\node[box, draw=red, line width=.5mm] at (0,2){};
\node[box, draw=red, line width=.5mm] at (2,2){};
\node[box, draw=red, line width=.5mm] at (2,0){};

 \draw[arw, ->, line width=0.5mm, color = blue](1,1)--(2,1);
 \draw[arw, ->, line width=0.5mm, color = blue](2,1)--(2,2);

\end{tikzpicture}

%% file: table_exp.tex
\begin{tabular}[t]{  | p{1.2cm} | p{1.8cm} | p{1.9cm} | p{.7cm}  |  p{.7cm}   | p{1.2cm} | p{1.8cm} |  p{1.5cm} |}
	\hline
	{\bf Type} & {\bf Quants} & {\bf Cases}  & {\bf $|S_P|$}  & {\bf $|S_Q|$}  & {\bf Z3} & {\bf Outcome }   & {\bf solve[s]}   \\ [0.5ex]
	\hline 
	\hline
	\multirow{6}{*}{\aesim} 
	 & \multirow{4}{*}{$\forall_\msmall~ \exists_\mbig$} 
	 & ABP & 11 & 14 & sat &  $|S_Q'|$=11 & 9.37  \\
	\cline{3-8}
	 &  & ABP$_\textsf{w/ bug}$ & 11 & 14 & unsat & - & 9.46  \\
	\cline{3-8}
	 &  & MM & 27 & 27 &  sat & $|S_Q'|$=27 & 67.74  \\
	\cline{3-8}
	 &  & MM$_\textsf{w/ bug}$ & 27 & 27 & unsat & - & 66.85  \\
	\cline{2-8}
	 &  \multirow{2}{*}{$\forall_\mbig~~~ \exists_\msmall$}  
	 & CBF & 15 & 9 & sat &  $|S_Q'|$=8 & 3.49 \\
	\cline{3-8}
	 &  & CBF$_\textsf{w/ bug}$ & 15 & 9 & unsat & - & 3.51   \\
	\hline
	\multirow{4}{*}{\easim} 
	& \multirow{2}{*}{$\exists_\msmall~ \forall_\mbig$}  
	& RP $3^3$ & 8 & 9 & sat &  $|S_P'|$=5 & 1.09  \\
	\cline{3-8}
	&  & RP $3^3$$_\textsf{no sol.}$ & 8 & 9 & unsat & - & 1.02 \\
	\cline{2-8}
	& \multirow{2}{*}{$\exists_\mbig~~~ \forall_\msmall$}  
	& GCW & 16 & 4 & sat & $|S_P'|$=8  & 3.36  \\
	\cline{3-8}
	&  & GCW$_\textsf{no sol.}$& 16 & 4 & unsat & - & 2.27  \\
	\hline
\end{tabular}

%% file: concl.tex
\section{Conclusion and Future Work}
\label{sec:concl}

We introduced efficient loop conditions for bounded model
checking of fragments of \HyperLTL.
We proved that considering only lasso-shaped traces is equivalent to
considering the entire trace set of the models, and proposed two
simulation-based algorithms $\easim$ and $\aesim$ to realize infinite
reasoning with finite exploration for \hltl formulas.
To handle non-determinism in the latter case, we combine the models
with prophecy automata to provide the (local) simulations with enough
information to select the right move for the inner $\exists$ path.
Our algorithms are implemented using \code{Z3py}.
We have evaluated the effectiveness and efficiency with successful
verification results for a rich set of input cases, which previous
bounded model checking approach would fail to prove.

As for future work, we are working on exploiting general prophecy
automata (beyond safety) in order to achieve full generality for the
$\forall\exists$ case.
The second direction is to handle asynchrony between the models in our
algorithm.
Even though model checking asynchronous variants of \HyperLTL is in
general undecidable~\cite{gutsfeld21automata,baumeister21temporal,bozzelli21asynchronous},
we would like to explore semi-algorithms and fragments with
decidability properties.
{\tacas Lastly, exploring how to handle infinite-state systems 
	with our framework by applying {\em abstraction} techniques is also another promising  
	future direction.} 

%% file: proofs.tex
\section{Proofs}

\newcounter{aux}

\subsection{Bounded Simulation}\label{sec:bounded.sim}
Let $K_1$ and $K_2$ be two Kripke structures over $AP$, and let $k\in \naturals$. 
The {\em bounded simulation problem} for $K_1,K_2$ and $k$ is to decide whether there exists a simulation relation from $K_1$ to $K_2$ that uses at most $k$ states of $K_2$ (note that in any case, all of the reachable states of $K_1$ must be used in such a simulation). 
We prove this problem to be \comp{NP-complete}.

\begin{theorem}
	The bounded simulation problem is \comp{NP-complete}.
\end{theorem}
\begin{proof}
	Let $K_1$ and $K_2$ be two Kripke structures with sets of states $Q_1$ and $Q_2$, respectively, and let $k\in \naturals$. 
	A nondeterministic algorithm which guesses a set $Q'_2\subseteq Q_2$ of at most $k$ states, and searches for a simulation from $K_1$ to $K_2$ reduced to $Q'_2$. Finding a simulation can be done in polynomial time, and so the problem is in \comp{NP}.
	
	We prove \comp{NP-hardness} by a reduction from the Vertex Cover problem. Given a directed graph $G=\tuple{V,E}$ where $|E|=m$ and where $V=\{v_1,\ldots v_n\}$, and $k\in\naturals$, we construct two Kripke structures $K_1$ and $K_2$, as follows.
	
	$K_1$ is composed of $m$ states, where for every $e\in E$ there is a state labeled $e$, and an additional initial state $q$ labeled $q$. The transitions are from $q$ to all edge states and vice versa. 
	
	$K_2$ is composed of $m$ states similarly labeled as the edge states of $K_1$, and additional $n$ states $v_1,\ldots v_n$  all labeled $q$, all initial. From every $v_i$ there are transitions to all edge states. From every edge state $(v_i,v_j)$ there are transitions to $v_i$ and $v_j$. 
	
	It is easy to see that $K_2$ can simulate $K_1$ using at most $m+k$ states iff $G$ has a vertex cover of size at most $k$.
\end{proof}